\documentclass[aps,prl,twocolumn,superscriptaddress]{revtex4}

\usepackage[a4paper, total={7in, 9in}]{geometry}

\usepackage{graphicx}
\usepackage{amsmath}
\usepackage{setspace}
\usepackage{color}
\usepackage{textcomp}
\usepackage{booktabs}
\usepackage[flushleft]{threeparttable}
\usepackage{comment}
\usepackage{amssymb}
\usepackage{amsmath}
\usepackage{float}
\usepackage{psfrag}
\usepackage{siunitx}
\usepackage{romannum}

\def\be{\begin{equation}}
\def\ee{\end{equation}}
\def\ba{\begin{eqnarray}}
\def\ea{\end{eqnarray}}

\begin{document}
\pagenumbering{arabic} 
\title{Marangoni instability of a drop in a stably stratified liquid}
\author{Yanshen Li}
\email{yanshen.li@utwente.nl}
\affiliation{Physics of Fluids group, Max-Planck Center Twente for Complex Fluid Dynamics, Department of Science and Technology, Mesa+ Institute, and 
J. M. Burgers Centre for Fluid Dynamics, University of Twente, P.O. Box 217, 7500 AE Enschede, The Netherlands}
\author{Christian Diddens}
\affiliation{Physics of Fluids group, Max-Planck Center Twente for Complex Fluid Dynamics, Department of Science and Technology, Mesa+ Institute, and 
J. M. Burgers Centre for Fluid Dynamics, University of Twente, P.O. Box 217, 7500 AE Enschede, The Netherlands}
\affiliation{Department of Mechanical Engineering, Eindhoven University of Technology, P.O. Box 513, 5600 MB Eindhoven, The Netherlands}
\author{Andrea Prosperetti}
\affiliation{Physics of Fluids group, Max-Planck Center Twente for Complex Fluid Dynamics, Department of Science and Technology, Mesa+ Institute, and 
J. M. Burgers Centre for Fluid Dynamics, University of Twente, P.O. Box 217, 7500 AE Enschede, The Netherlands}
\affiliation{Department of Mechanical Engineering, University of Houston, Texas 77204-4006, USA}
\author{Detlef Lohse}
\email{d.lohse@utwente.nl}
\affiliation{Physics of Fluids group, Max-Planck Center Twente for Complex Fluid Dynamics, Department of Science and Technology, Mesa+ Institute, and 
J. M. Burgers Centre for Fluid Dynamics, University of Twente, P.O. Box 217, 7500 AE Enschede, The Netherlands}
\affiliation{Max Planck Institute for Dynamics and Self-Organization, Am Fa\ss berg 17, 37077 G\"ottingen, Germany}
\begin{abstract} 
Marangoni instabilities can emerge when a liquid interface is subjected to a concentration or temperature gradient. It is generally believed that for these instabilities bulk effects like buoyancy are negligible as compared to interfacial forces, especially on small scales. Consequently, the effect of a stable stratification on the Marangoni instability has hitherto been ignored. Here we report, for an immiscible drop immersed in a stably stratified ethanol-water mixture, a new type of oscillatory solutal Marangoni instability which is triggered once the stratification has reached a critical value. We experimentally explore the parameter space spanned by the stratification strength and the drop size and theoretically explain the observed crossover from levitating to bouncing by balancing the advection and diffusion around the drop. Finally, the effect of the stable stratification on the Marangoni instability is surprisingly amplified in confined geometries, leading to an earlier onset.
\end{abstract}

\maketitle
\newpage

A concentration or temperature gradient applied to an interface can induce a Marangoni instability of the motionless state, resulting in a steady convection. Similarly, the steady state Marangoni convection can undergo another instability, leading to an oscillatory motion \cite{rednikov1998two}. Since the first quantitative analysis in 1958 \cite{pearson1958convection}, Marangoni instabilities have been studied extensively due to their relevance for liquid extraction \cite{sternling1959interfacial, groothuis1960influence, rother1999effect, berejnov2002spontaneous, jain2011recent}, coating techniques \cite{pearson1958convection, yarin1995surface, demekhin2006suppressing}, metal processing \cite{gupta1992pore, ratke2005theoretical, zhang2006indirect} and crystal growth \cite{schwabe1978experiments, schwabe1979some, chang1979thermocapillary, chun1979experiments, schwabe1982studies, preisser1983steady, kamotani1984oscillatory}, etc. Marangoni instabilities are also the main mechanism to drive the self-propulsion of active drops \cite{rednikov1994active, rednikov1994drop, herminghaus2014interfacial, yoshinaga2014spontaneous, ryazantsev2017thermo, maass2016swimming, morozov2019self}, which have attracted lots of recent interest. Such drops are an example of the rich physicochemical hydrodynamics of droplets far from equilibrium \cite{lohse2020physicochemical} which are very relevant for food processing \cite{degner2013influence, degner2014factors} and modelling biological systems \cite{maass2016swimming}, etc. 

Depending on the application, Marangoni instabilities have been investigated in different configurations, such as a horizontal interface between two fluid layers \cite{pearson1958convection, sternling1959interfacial, reichenbach1981linear, takashima1981surface, levchenko1981instability, nepomnyashchii1983thermocapillary, chu1988sustained, chu1989transverse, hennenberg1992transverse, rednikov1998two}, the surface of a falling film on a tilted plate \cite{nepomnyashchy1976wavy, chang1994wave, kliakhandler1997viscous, miladinova2005effects, demekhin2006suppressing}, a vertical interface of a liquid column \cite{schwabe1978experiments, schwabe1979some, chang1979thermocapillary, chun1979experiments, schwabe1982studies, preisser1983steady, kamotani1984oscillatory}, and for drops submerged in a solution \cite{rednikov1994active, rednikov1994drop, herminghaus2014interfacial, yoshinaga2014spontaneous, ryazantsev2017thermo, maass2016swimming, morozov2019self, thanasukarn2004impact, ghosh2008factors, degner2013influence, degner2014factors, dedovets2018five}, etc. In many of these situations, these systems are subjected to a stabilizing temperature/concentration gradient \cite{takashima1981surface, levchenko1981instability, demekhin2006suppressing, schwabe1978experiments, schwabe1979some, chang1979thermocapillary, chun1979experiments, schwabe1982studies, preisser1983steady, kamotani1984oscillatory, chu1988sustained, chu1989transverse}, which induces a continuously stable density stratification.
However, except for a few cases for the horizontal interface configuration \cite{welander1964convective, wierschem2000internal, rednikov2000rayleigh}, the effect of such a stable density stratification on Marangoni convection has always been ignored, due to the generally accepted view that on small scales bulk effects like buoyancy are negligible \cite{nepomnyashchy2012interfacial}. Here we report, for an immiscible drop immersed in an ethanol-water mixture, that the stable stratification could actually trigger an oscillatory instability once it is above a critical value.  Surprisingly, this critical value will decrease in a confined geometry, implying that the effect of the stable stratification is actually \textit{amplified} on small scales. Our findings demonstrate that stable stratification can strongly affect Marangoni convection and ask for further studies in related geometries.

\begin{figure*}[ht!]
\begin{center} 
\includegraphics[width=0.88\textwidth]{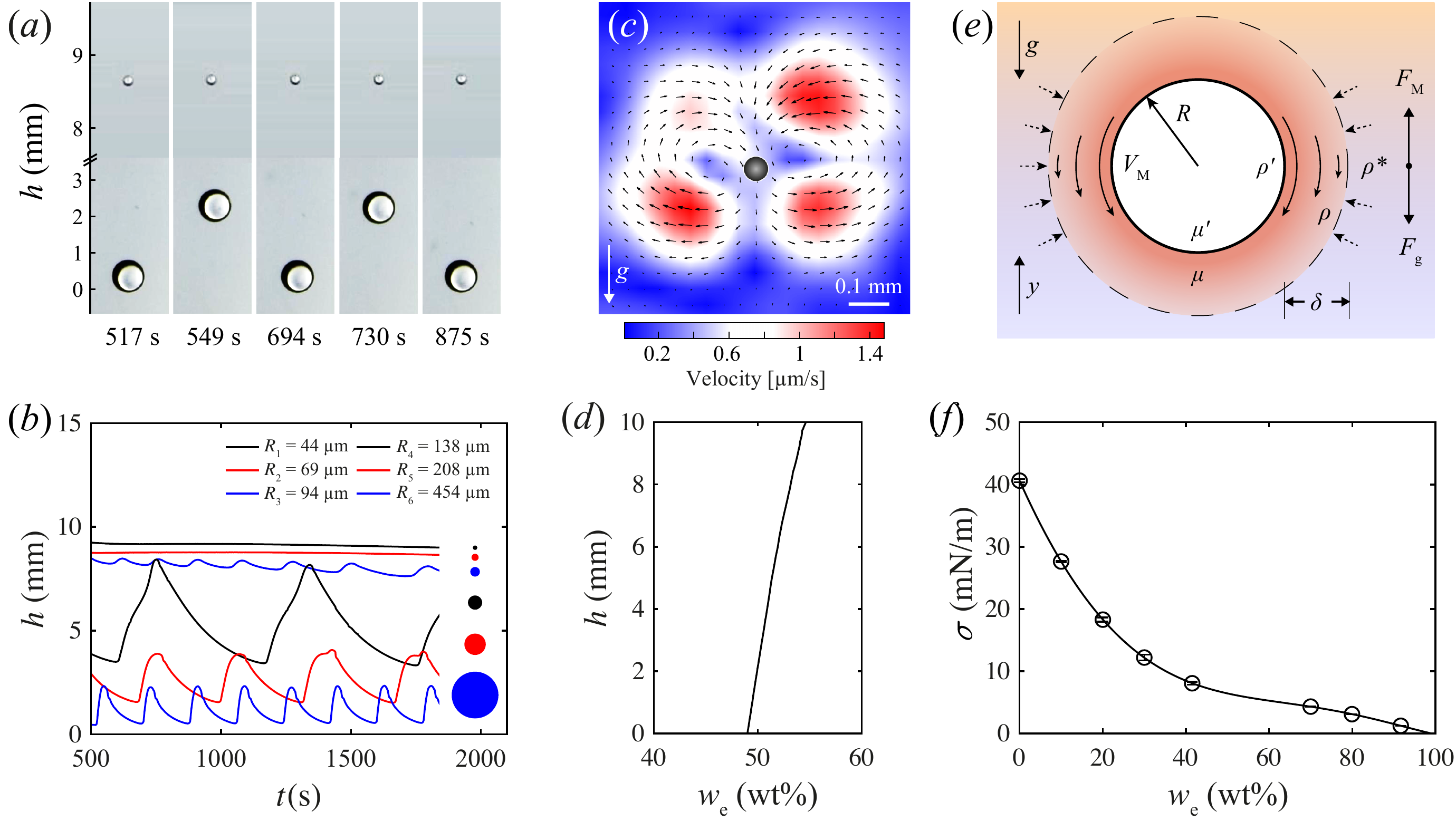}
\caption{\setstretch{1}Bouncing and levitating drops in a linearly and stably stratified mixture of ethanol (lighter) and water (heavier). (a) Snapshots of two 5 cSt silicone oil drops at the given time after they were released in the mixture. The larger drop bounces at $h<\SI{3}{mm}$, while the smaller drop is levitating at a higher position $h\approx\SI{8.7}{mm}$. The snapshots are taken from one experiment with two drops. To better show them, the upper/lower half of the snapshots are shown with different scales. (b) Drop's height $h$ as functions of time $t$ for different drop radii $R$ after the initial sinking period. $h=0$ is the position where the density of the drop equals that of the mixture. The filled circles represent the relative size of the drops. (c) Flow field around a levitating drop ($R=31\pm\SI{1}{\micro\meter}$) measured by PIV, in a mixture with $\mathrm{d} w_\mathrm{e}/\mathrm{d} y\approx\SI{5}{m^{-1}}$. The resolution is not high enough to resolve the velocity close to the drop's surface. (d) The ethanol weight fraction $w_\mathrm{e}$ at the corresponding height, with $\mathrm{d} w_\mathrm{e}/\mathrm{d} y\approx\SI{5}{m^{-1}}$. (e) A sketch of the levitating drop (with radius $R$ and density $\rho^\prime$) and the ethanol concentration around it. Deeper red means higher ethanol concentration. The shaded ring inside the dashed circle represents the kinematic boundary layer with thickness $\delta$, set by the Marangoni velocity $V_\mathrm{M}$. The ethanol concentration inside this layer is enhanced \& homogenized by Maragnoni advection bringing down the ethanol rich liquid. The Marangoni flow is represented by the solid arrows. Dashed arrows represent diffusion across this layer. $\rho$ is the representative density inside this layer, and $\rho^*$ is the undisturbed density in the far field. $\mu$ and $\mu^\prime$ are the viscosities of the mixture and the drop, respectively. (f) Interfacial tension $\sigma(w_\mathrm{e})$ between \SI{5}{cSt} silicone oil and the ethanol-water mixture. Each point is an average of six measurements and the error bar is the standard deviation. The solid line is a polynomial fit to the data points.}
\label{fig:1}
\end{center}
\end{figure*}

To determine the onset of the Marangoni instability, we experimentally explore the parameter space spanned by the concentration gradient and the drop radius $R$. Using the double-bucket method \cite{oster1965density}, linearly stratified liquid mixtures are prepared in a cubic glass container (Hellma, 704.001-OG, Germany) with inner width of $L=\SI{30}{mm}$ filled to different depth, depending on the degree of stratification. The ethanol weight fraction $w_\mathrm{e}$ at each height is measured by laser deflection \cite{lin2013one, li2019bouncing}, from which the gradient of ethanol weight fraction $\mathrm{d} w_\mathrm{e}/\mathrm{d} y$ is calculated. The concentration gradient $\mathrm{d} w_\mathrm{e}/\mathrm{d} y$ is varied from $\sim\SI{3}{m^{-1}}$ to $\sim\SI{130}{m^{-1}}$, corresponding to density gradients ranging from \SI{-480}{kg/m^4} to \SI{-4200}{kg/m^4}. \SI{5}{cSt} Silicone oil (Sigma-Aldrich, Germany) is injected through a thin needle (with outer-diameter \SI{0.515}{mm}) to generate drops of different radii $R$. The drops are released from the top of the stratified mixtures, and their trajectories are recorded by a sideview camera. During the measurements, only one single drop exists in the container at a time. The silicone oil has density $\rho^\prime=\SI{913}{kg/m^3}$ and viscosity $\mu^\prime=\SI{4.6}{\milli\pascal\cdot\second}$.

Two typical behaviors are observed after the initial sinking phase. See Fig. \ref{fig:1}(a) for the successive snapshots of two silicone oil drops in a mixture with $\mathrm{d} w_\mathrm{e}/\mathrm{d} y\approx\SI{5}{m^{-1}}$: While a smaller drop ($R=69\pm\SI{2}{\micro\meter}$) stays at a fixed position around $h\approx\SI{8.7}{mm}$, a larger drop ($R=454\pm\SI{2}{\micro\meter}$) bounces continuously in the range $\SI{0}{mm}<h<\SI{3}{mm}$. Here $h=0$ marks the position where the density of the oil ($\rho^\prime=\SI{913}{kg/m^3}$) equals that of the mixture (at $w_\mathrm{e}\approx\SI{49}{\%}$). The drop's position $h(t)$ as a function of time $t$ in the same stratified liquid and the ethanol weight fraction $w_\mathrm{e}$ at the corresponding height are respectively shown in Fig. \ref{fig:1}(b) and (d). The smallest drop ($R_1\approx\SI{44}{\micro\meter}$) is levitating at $h\approx\SI{9.1}{mm}$. As the drop size increases, it levitates at a lower position, until above a critical radius $R_\mathrm{cr}$ it starts to bounce instead of levitating. If its size is further increased, the drop bounces around a lower position (but still with $h>0$). 

The smaller drops are able to levitate above the density matched position $h=0$ against gravity because of a stable Marangoni flow around it, as shown in Fig. \ref{fig:1}(c). The flow field is obtained by PIV measurements for a drop levitating in the gradient $\mathrm{d} w_\mathrm{e}/\mathrm{d} y\approx\SI{5}{m^{-1}}$. The interfacial tension of the drop $\sigma$ decreases with increasing ethanol concentration of the mixture $w_\mathrm{e}$, as shown in Fig. \ref{fig:1}(f). This interfacial tension gradient at the drop's surface pulls liquid downwards, generating a viscous force acting against gravity, which levitates the drop. When the drop becomes large enough, however, the equilibrium becomes oscillatory, and the drop starts to bounce between two different levels. Thus, the transition from a levitating drop to a bouncing one signals the onset of the instability. 

\begin{figure}[t!] 
\begin{center} 
\includegraphics[width=0.35\textwidth]{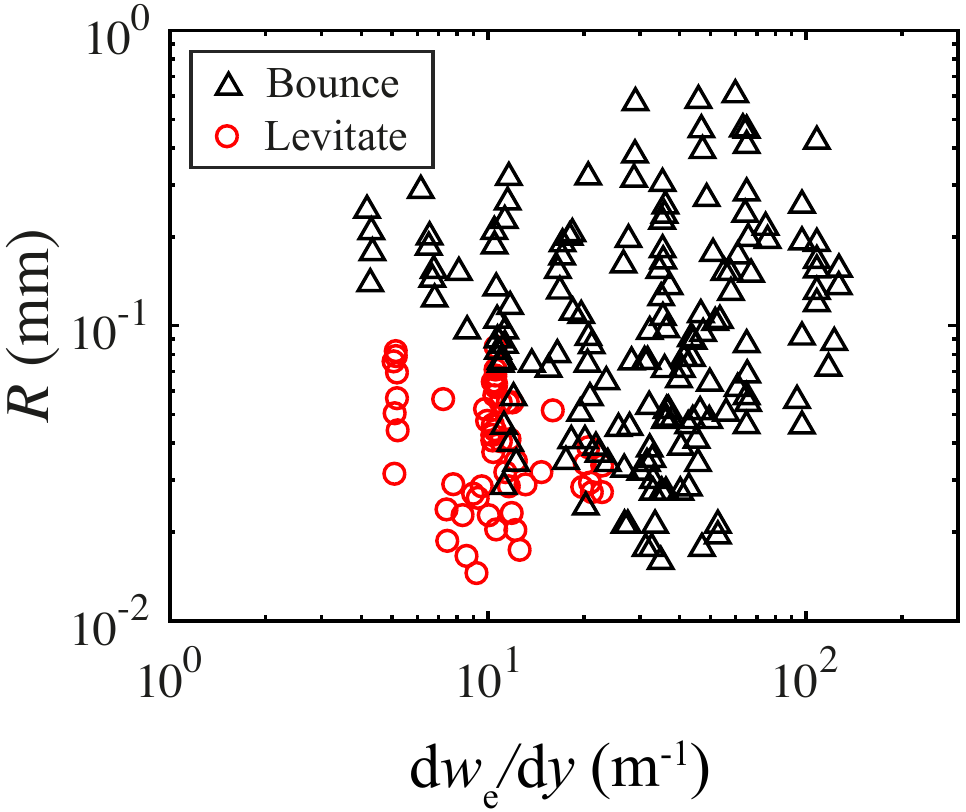}
\caption{\setstretch{1}Phase diagram of the levitating \& bouncing drops in the parameter space of drop radius $R$ vs. concentration gradient $\mathrm{d}w_\mathrm{e}/\mathrm{d}y$. Black triangles stand for bouncing drops, red circles for levitating ones. Measurement errors in the $x$ direction are comparable with the size of the symbols.}
\label{fig:2}
\end{center}
\end{figure}

While exploring the parameter space, we use an easily distinguishable criterion to determine whether a drop is bouncing: If the drop's bouncing amplitude $h_\mathrm{A}$ is larger than its radius $R$, then the drop is considered to be bouncing (see Supplemental Material for more details). The results are shown in Fig. \ref{fig:2}. Surprisingly, while for weak gradients (like for $\mathrm{d}w_\mathrm{e}/\mathrm{d}y\approx\SI{10}{m^{-1}}$) there is a critical radius $R_\mathrm{cr}$ ($\approx\SI{80}{\micro\meter}$) above which the Marangoni flow becomes unstable, the Marangoni flow is always unstable for stronger gradients $\mathrm{d}w_\mathrm{e}/\mathrm{d}y>\SI{23}{m^{-1}}$ in all performed experiments. Note that for larger drops ($R>\SI{0.1}{mm}$), we could not explore the full parameter space for $\mathrm{d}w_\mathrm{e}/\mathrm{d}y<\SI{3}{m^{-1}}$ since it would require an unrealistically large container. 

To get a better understanding of the onset of this Marangoni instability, the key is to understand the coupling between the Marangoni flow and the concentration field: The Marangoni flow is induced by the ethanol (solute) concentration gradient around the drop, which is subjected to change by advection (caused by the Marangoni flow itself) and diffusion, see the sketch in Fig. \ref{fig:1}(e). The Marangoni flow tends to homogenize the ethanol concentration around the drop, thus weakening the Marangoni flow force and thus itself. At the same time, diffusion acts to restore the ethanol gradient in the vicinity of the drop to its undisturbed value, i.e., the value it takes in the far field. This competition between advection and diffusion around the drop determines whether the Marangoni flow is stable or not. Furthermore, once it becomes unstable, a temporarily strong Marangoni flow homogenizes the concentration field around the drop, consequently weakening itself. Later the Marangoni flow restarts once diffusion has restored the concentration field around the drop, so that the flow is oscillatory and leads to the continuous bouncing of the drop. 

The liquid layer whose concentration is affected by the Marangoni advection is effectively the {Marangoni flow} boundary layer with thickness $\delta$ 
(see Fig. \ref{fig:1}(e)). The time scale for advection to change the concentration in this layer is the time needed for the Marangoni flow to bring down the ethanol-rich liquid from the top: $\tau_\mathrm{a}\sim R/V_\mathrm{M}$, where $V_\mathrm{M}$ is the Marangoni flow velocity at the equator of the drop. For the drop in the concentration gradient it holds \cite{young1959motion} (see Supplementary Material) $V_\mathrm{M}\sim-\mathrm{d}\sigma/\mathrm{d}y\cdot R/(\mu+\mu^\prime)$. The time scale for diffusion to restore the concentration across this layer is $\tau_\mathrm{d}\sim \delta^2/D$, where $D$ is the diffusivity of ethanol in water. The flow will become unstable when advection is faster than diffusion, $\tau_\mathrm{a}<\tau_\mathrm{d}$.
Substituting the two time scales into this relation, we obtain ${V_\mathrm{M}R}/{D}>{R^2}/{\delta^2}$. The left hand side has the form of a P\'eclet number, which is the ratio between advection and diffusion, and which in problems of this type is referred to as the Marangoni number
\begin{equation}
Ma=\frac{V_\mathrm{M}R}{D}=-\frac{\mathrm{d}\sigma}{\mathrm{d}w_\mathrm{e}}\frac{\mathrm{d}w_\mathrm{e}}{\mathrm{d}y}R^2\cdot\frac{1}{(\mu+\mu^\prime)D}, 
\label{eq:1}
\end{equation}
where we have used above expression for $V_\mathrm{M}$ with an equal sign and where $\mathrm{d}\sigma/\mathrm{d}w_\mathrm{e}$ is a material property (see Fig. \ref{fig:1}(f)) and $\mathrm{d}w_\mathrm{e}/\mathrm{d}y$ the undisturbed ethanol gradient of the mixture. The instability criterion thus is 
\begin{equation}
Ma>{R^2}/{\delta^2}.
\label{eq:2}
\end{equation}

The liquid within the boundary layer is lighter than its surroundings as it is entrained from the top, and it is held in place by the Marangoni induced viscous stress against buoyancy:
\begin{equation}
\mu\frac{V_\mathrm{M}}{\delta^2}\sim g\Delta\rho ,
\label{eq:3}
\end{equation}
where $\Delta\rho=\rho^*-\rho$ is the density difference between the liquid inside and outside of the kinematic boundary layer, see Fig. \ref{fig:1}(e). The lighter liquid is brought down by the Marangoni flow along the drop's surface, so $\Delta\rho\sim -R\cdot\mathrm{d}\rho/\mathrm{d}y$. Cancelling $\delta$ from Eqs.(\ref{eq:2})\&(\ref{eq:3}),  we obtain the instability criterion
\begin{equation}
Ma/{Ra}^{1/2}> c,
\label{eq:4}
\end{equation}
where 
\begin{equation}
Ra=-\frac{\mathrm{d}\rho}{\mathrm{d}y}\cdot \frac{gR^4}{\mu D}
\end{equation}
is the Rayleigh number for characteristic length $R$ and $c$ is a constant to be determined. 

To calculate the Marangoni and Rayleigh numbers, ethanol weight fractions at the positions where the drops levitate are used to obtain the viscosity $\mu$, diffusivity $D$ and the interfacial tension $\sigma$ (see Supplemental Material for the concentration dependence of $\mu$ and $D$). In the following, for bouncing drops, we use values corresponding to their lowest position.

\begin{figure}[t!]
\begin{center} 
\includegraphics[width=0.35\textwidth]{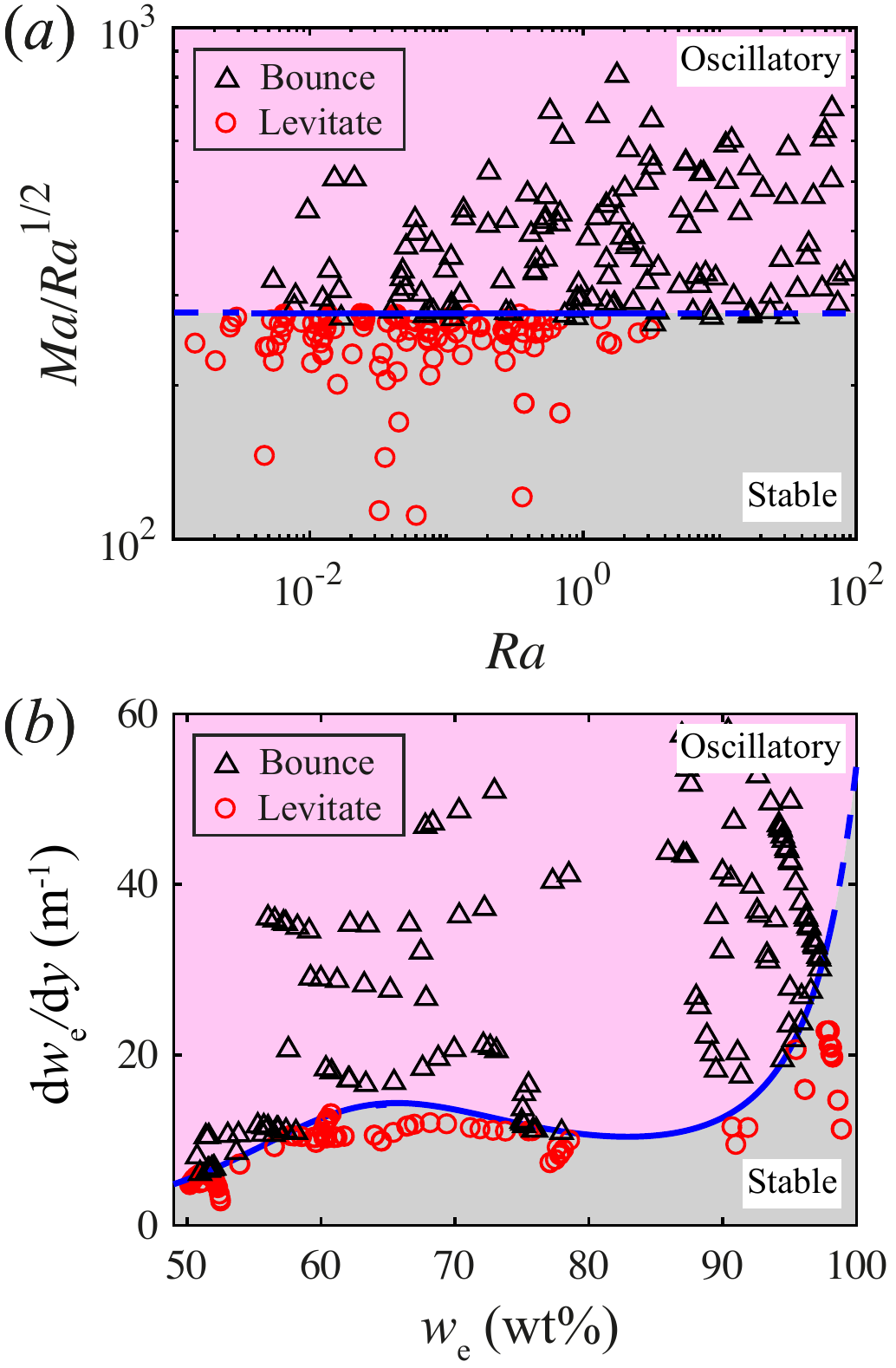}
\caption{\setstretch{1}(a) Phase diagram replotted in dimensionless numbers: $Ma/Ra^{1/2}$ vs. $Ra$. Black triangles stand for bouncing drops, red circles for levitating ones. The blue line is the instability threshold $(Ma/{Ra}^{1/2})_\mathrm{cr}=275$, above which the flow is oscillatory and all drops bounce. The blue solid line (in the range $6\times10^{-3}<Ra<3$) is confirmed by experiments. Measurement errors in the $y$ direction are comparable with the size of the symbols. (b) Phase diagram replotted with $\mathrm{d}w_\mathrm{e}/\mathrm{d}y$ vs. $w_\mathrm{e}$, where $w_\mathrm{e}$ is the ethanol weight fraction at the levitation height. The blue curve is calculated from Eq.(\ref{eq:6}) with $c=275$. The dashed blue line in the range $w_\mathrm{e} >\SI{98}{wt\%}$ ($w_\mathrm{e} < \SI{50}{wt\%}$) corresponds to $Ra< 6\times10^{-3}$ ($Ra>3$). Measurement errors are comparable with the size of the symbols.}
\label{fig:3}
\end{center}
\end{figure}

The phase diagram shown in Fig. \ref{fig:2} is replotted with $Ma/Ra^{1/2}$ vs. $Ra$ in Fig. \ref{fig:3}(a). It clearly shows that there is indeed a critical value $(Ma/{Ra}^{1/2})_\mathrm{cr}$ above which the drop will always bounce, and the instability threshold $c$ in Eq.(\ref{eq:4}) is measured to be $c=275\pm10$ in the range $6\times10^{-3}\lesssim Ra\lesssim3$. We cannot carry out experiments for $Ra<6\times10^{-3}$ because drops with $R<\SI{20}{\micro\meter}$ are too small to observe. For experiments in the range $Ra>3$, the finite size of the container comes into play. However, we speculate that $c\approx275$ still holds for $Ra > 3$ as long as the container is large enough. The existing data on bouncing are consistent with this value.

We now express our stability criterion Eq.(\ref{eq:4}) in dimensional quantities by substituting the definition of $Ma$ and $Ra$ to obtain:
\begin{equation}
\left(\frac{\mathrm{d}w_\mathrm{e}}{\mathrm{d}y}\right)_\mathrm{cr}=c^2\left(\mu+\mu^\prime\right)^2\cdot\frac{gD}{\mu}\frac{\mathrm{d}\rho}{\mathrm{d}\sigma}\frac{\mathrm{d}w_\mathrm{e}}{\mathrm{d}\sigma}.
\label{eq:6}
\end{equation}
Eq.(\ref{eq:6}) actually predicts a critical concentration gradient above which the equilibrium is unstable. Note that remarkably the drop radius $R$ does not enter into this equation. All the fluid properties $\mu$, $D$, $\mathrm{d}\rho/\mathrm{d}\sigma$ and $\mathrm{d}w_\mathrm{e}/\mathrm{d}\sigma$ depend on $w_\mathrm{e}$ -- the ethanol weight fraction at the levitation height. Thus the critical gradient $(\mathrm{d}w_\mathrm{e}/\mathrm{d}y)_\mathrm{cr}$ as a function of $w_\mathrm{e}$ is shown in Fig. \ref{fig:3}(b) as the blue curve. The data shown in Fig. \ref{fig:2} are also replotted in Fig. \ref{fig:3}(b). As can be seen, the blue curve as predicted by Eq.(\ref{eq:6}) nicely separates the levitating drops and the bouncing ones. The dashed blue line in the range $w_\mathrm{e} >\SI{98}{wt\%}$ ($w_\mathrm{e} < \SI{50}{wt\%}$) corresponds to $Ra< 6\times10^{-3}$ ($Ra>3$), i.e., the region in which we could not perform experiments. 

The above results are all obtained in a large enough container. We will now discuss the effect of a geometrical confinement, i.e., the dependence of our findings on the container size $L$. Let $\mathcal{L}$ denote the maximum extent of the flow field induced by the drop. Then $\mathcal{L}>L$ means that the flow is confined. In the case of no confinement, i.e., $\mathcal{L}<L$, the liquid in the far field is not disturbed by the Marangoni flow, so that the density in the far field is maintained at $\rho^*$ (see Fig. \ref{fig:1}(e)). However, when the flow is confined, i.e., $\mathcal{L}>L$, the liquid close to the side wall is affected by the Marangoni flow. In such a situation, because the liquid is pulled down in the center by the drop, the liquid close to the wall will be pushed up due to mass conservation. This effectively increases the density $\rho^*$. Consequently, the density difference $\Delta\rho=\rho^*-\rho$ is increased, which means that the effect of buoyancy is amplified. According to Eqs. (\ref{eq:3})\&(\ref{eq:4}), the instability threshold $c$ will thus decrease. Either decreasing the container size $L$ or increasing $\mathcal{L}$ both leads to a stronger confinement effect. Since for stable stratifications $\mathcal{L}\sim\left(-{\mathrm{d}\rho}/{\mathrm{d}y}\cdot{\mu D}/{g}\right)^{-1/4}$ \cite{phillips1970flows, wunsch1970oceanic}, one can thus also increase the confinement effect by using very weak stratifications. We have performed experiments for weaker gradients $\mathrm{d}w_\mathrm{e}/\mathrm{d}y<\SI{3}{m^{-1}}$ and also in a larger container to confirm the effect of the confinement. Indeed, for $\mathrm{d}w_\mathrm{e}/\mathrm{d}y\approx\SI{2}{m^{-1}}$, a cubic container with $L=\SI{50}{mm}$ is already not large enough, and the instability threshold is reduced to $c\approx172$. A smaller container ($L=\SI{30}{mm}$) further decreases the threshold to $c\approx157$. For $\mathrm{d}w_\mathrm{e}/\mathrm{d}y\approx\SI{1}{m^{-1}}$, the geometry is more confined, and the threshold is further decreased to $c\approx122$ in the larger container and even to $c\approx117$ in the smaller one. The fact that a weaker stratification leads to a more amplified effect of buoyancy demonstrates that the stable stratification is very relevant for the Marangoni instability, in particular on small scales where the confinement is more relevant.

In conclusion, we have discovered a new type of oscillatory Marangoni instability for an immiscible drop immersed in a stably stratified ethanol-water mixture. The commonly ignored stable density stratification induced by the concentration gradient is vital in triggering this instability. Its onset is indicated by the transition from a levitating drop to a bouncing one. By experimentally exploring the parameter space spanned by the concentration gradient $\mathrm{d}w_\mathrm{e}/\mathrm{d}y$ and the drop radius $R$, the instability is found to be determined by the balance between the advection and diffusion through the kinetic boundary layer set by the Marangoni flow. This yields a critical concentration gradient as the instability criterion. Remarkably, the critical gradient is decreased in a confined geometry, i.e., the effect of the stable stratification is amplified on small confined scales. Our findings indicate that the stable stratification induced by the corresponding concentration gradient is very relevant, especially in confined geometries, and should be further explored in other geometries.

Our results for solutal Marangoni flows can also be extended to thermal Marangoni flows. We found that a stabilizing temperature gradient as low as $\SI{3}{K/mm}$ can trigger a similar oscillatory instability on a bubble immersed in water. Such low temperature gradient is smaller than what is occurring in various applications \cite{schwabe1978experiments, ratke2006destabilisation, dedovets2018five}, where a stabilizing temperature gradient can easily go beyond \SI{10}{K/mm}.

We thank Chao Sun and Vatsal Sanjay for valuable discussions. We acknowledge support from the Netherlands Center for Multiscale Catalytic Energy Conversion (MCEC), an NWO Gravitation programme funded by the Ministry of Education, Culture and Science of the government of Netherlands, and the ERC-Advanced Grant Diffusive Droplet Dynamics (DDD) with Project No. 740479.


\end{document}